\documentclass[hyphens]{article}

\usepackage[utf8]{inputenc} 
\usepackage[T1]{fontenc}    
\usepackage{hyperref}       

\usepackage{booktabs}       
\usepackage{amsfonts}       
\usepackage{nicefrac}       
\usepackage{microtype}      
\usepackage{xcolor}         
\usepackage{amsmath}
\usepackage{graphicx}
\usepackage{multirow}
\usepackage[preprint]{neurips_2026}

\title{Multi Codec Discrete Diffusion Model for\\Text Guided Speech Inpainting and Editing}

\author{%
  Iftach Shoham\textsuperscript{*,1}
  \And
  Tali Dror\textsuperscript{*,1}
  \AND
  Oren Gal\textsuperscript{2}
  \And
  Haim Permuter\textsuperscript{1}
  \And
  Gilad Katz\textsuperscript{1}
  \And
  Eliya Nachmani\textsuperscript{1}
  \\[1.2ex]
  \textsuperscript{1}Ben-Gurion University of the Negev
  \qquad
  \textsuperscript{2}University of Haifa
  \\[0.4ex]
  \textsuperscript{*}Equal contribution
}

\begin{document}

\maketitle

\begin{abstract}
Speech recordings often contain missing, corrupted, or incorrect regions that must be reconstructed or modified without re-synthesizing the entire utterance. Speech inpainting restores missing segments, whereas speech editing replaces spoken content according to an edited transcript. Both tasks require the generated speech to express the intended words while remaining consistent with the surrounding speaker identity, prosody, timing, and recording conditions. Discrete diffusion is particularly well suited to these tasks because it can iteratively refine masked tokens while jointly conditioning on both left and right acoustic context. We introduce SIEDD, a discrete diffusion framework for text-guided speech inpainting and editing over hierarchical codec tokens. Its core architecture, HiCoDD, follows the RVQ generation order by representing previously generated codebooks as clean, committed acoustic context and applying diffusion only to the current refinement codebook. This separation enables leakage-free joint training while matching sequential coarse-to-fine inference. The model further combines phoneme-level conditioning, span-localized classifier-free guidance, and duration prediction to support both fixed-duration inpainting and variable-duration text edits. On the RealEdit benchmark, SIEDD achieves the best overall speech-editing performance among the evaluated methods. It also outperforms the evaluated autoregressive baselines across all speech-inpainting settings, on both single and multiple gaps. These results demonstrate that explicitly modeling the codec hierarchy substantially improves context-preserving speech reconstruction and editing. See our full code\footnote{\url{https://github.com/iftachShoham/SIEDD}}.
\end{abstract}

\section{Introduction}
\label{introduction}

The reconstruction of missing or corrupted regions in an audio signal is commonly referred to as \textit{audio inpainting}~\cite{le2008computational,le2011computational}. Speech inpainting is a particularly challenging instance of this problem because the reconstructed segment must be both acoustically plausible and linguistically meaningful. It must also remain consistent with the surrounding speaker identity, prosody, rhythm, coarticulation, and recording environment. These requirements arise in applications such as repairing damaged recordings, removing localized noise or artifacts, recovering packet losses, and restoring omitted speech. Text-guided speech editing extends this problem from restoration to intentional modification. Given an existing utterance and an edited transcript, the goal is to insert, delete, or replace spoken content without re-synthesizing the entire recording. The generated segment must express the requested text while blending naturally with the unmodified speech at its boundaries.

Traditional inpainting methods relied on signal-processing assumptions such as local stationarity, autoregressive (AR) prediction~\cite{adler2011audio}, sparse representations~\cite{marafioti2020gacela}, or linear predictive coding~\cite{adler2011audio}. These methods can be effective for very short gaps, but they often fail when the missing region spans phonemes, syllables, or words, where reconstruction requires longer-range linguistic and acoustic context. Recent work has therefore shifted toward speech-aware generative modeling.

Text-informed and text-conditioned speech inpainting methods use the transcript to guide reconstruction of the missing region~\cite{prablanc2016text,borsos2022speechpainter}. Closely related speech editing systems extend this setting to intentional replacement of spoken content according to an edited transcript~\cite{tan2021editspeech,wang2022campnet}.

More recent approaches use stronger generative backbones, including diffusion, flow matching, and neural codec language models, enabling higher-quality infilling and zero-shot speech editing~\cite{le2023voicebox,wang2025ssr,peng2024voicecraft,huang2024instructspeech}. However, speech editing is not naturally a left-to-right generation problem. The missing span is jointly constrained by the requested text and by the acoustic context on both sides. AR codec models can condition on both boundaries, but still advance through the acoustic sequence in a fixed causal order. Earlier temporal positions must be generated and committed before later ones and cannot subsequently be revised. Errors in linguistic content, duration, or prosody may therefore propagate through the edited span.

Discrete diffusion offers a natural alternative by formulating editing as reverse-time masked reconstruction. It can jointly generate and repeatedly refine temporal positions under both left and right acoustic context, without imposing a fixed left-to-right order inside the missing span. This formulation also naturally supports the simultaneous reconstruction of multiple disjoint gaps.

Applying discrete diffusion to neural codec tokens, however, introduces a separate challenge. Residual vector quantization (RVQ) codebooks (CBs) are not interchangeable: lower CBs establish coarse linguistic and acoustic structure, while higher CBs encode residual refinements conditioned on the levels below them. A naive diffusion model that flattens or jointly denoises all CBs treats these levels symmetrically and neglects the hierarchy under which the codec representation was constructed.

We address both limitations with \textit{Speech Inpainting and Editing via Discrete Diffusion} (SIEDD), a text-conditioned framework that retains diffusion's joint refinement across time while explicitly preserving the coarse-to-fine dependencies across RVQ CBs.

Our contributions are threefold:
\begin{itemize}
    \item \textbf{Hierarchy-aware codec diffusion.} We introduce HiCoDD, which jointly denoises temporal positions within each CB while generating CBs in their inherent coarse-to-fine order. A clean-CB encoder conditions the noised-CB decoder on previously committed RVQ levels, preserving cross-CB dependencies without imposing AR generation across time.

     \item \textbf{Contrastive guidance for text-conditioned inpainting and editing.} We adapt classifier-free guidance (CFG) for categorical score prediction, combining conditional and negative concrete scores in log-ratio space. The negative branch localizes the contrast to the edited span, so it suppresses the target phonetic content while preserving the surrounding transcript. 

    \item \textbf{Strong speech-inpainting and editing performance.} On the RealEdit benchmark, SIEDD achieves the best overall speech-editing performance among the evaluated methods, including the lowest WER and MCD and the highest speaker similarity. It also outperforms the evaluated AR baselines across all speech-inpainting settings.
\end{itemize}

Together, these contributions establish a hierarchy-aware discrete-diffusion framework for speech inpainting and editing. SIEDD preserves the coarse-to-fine structure of RVQ codecs, conditioning generation on both the requested linguistic content and the surrounding acoustic context.

\section{Related Work}
\label{related_work}

\subsection{Discrete Diffusion for Language, Speech, and Audio}
\label{subsec:discrete_diffusion_related}

Discrete diffusion models (DDM) provide an alternative to AR generation for categorical sequences. Following D3PM~\cite{austin2021structured}, subsequent work introduced diffusion-based language models such as DiffusionBERT~\cite{he2023diffusionbert}, score-entropy discrete diffusion (SEDD)~\cite{lou2310discrete}, masked diffusion language models~\cite{sahoo2024simple}, and large-scale diffusion language models such as LLaDA~\cite{nie2026large}. Related advances include discrete flow matching~\cite{gat2024discrete} and remasking strategies for iterative token correction~\cite{wang2026remasking}. By replacing fixed left-to-right decoding with iterative denoising, these methods enable parallel prediction and flexible quality-speed trade-offs. Discrete diffusion has also been extended to speech and audio, including text-to-sound generation~\cite{yang2023diffsound}, speech recognition~\cite{baas2022transfusion}, audio inpainting~\cite{dror2026token}, and generative modeling of text-aligned speech tokens~\cite{ku2026discrete}.

\subsection{Speech Inpainting and Editing}
\label{subsec:speech_inpainting_editing}

Audio inpainting restores missing or corrupted regions \cite{le2008computational} and encompasses related formulations such as interpolation \cite{marafioti2020gacela}, extrapolation and imputation \cite{lieb2018audio}, and waveform substitution \cite{adler2011audio}. Classical approaches, generally limited to short gaps, rely on AR prediction \cite{adler2011audio}, sparse STFT or Gabor representations \cite{lieb2018audio,adler2011audio}, adaptive dictionaries \cite{taubock2020dictionary}, NMF \cite{mokry2023algorithms}, sinusoidal models \cite{adler2011audio}, or graph regularization \cite{perraudin2018inpainting}. Speech inpainting additionally requires preserving linguistic content, speaker identity, prosody, and recording conditions, motivating text-informed synthesis and voice conversion \cite{prablanc2016text}, text-conditioned generation \cite{borsos2022speechpainter}, audio-visual conditioning \cite{montesinos2023speech}, and diffusion models incorporating text, phonemes, or self-supervised speech representations \cite{yang2024usee,moradi2025transient}.

Text-guided speech editing extends inpainting to intentional deletion, insertion, or replacement while maintaining consistency with the surrounding audio. Early systems combined synthesis, voice conversion, and waveform stitching \cite{jin2017voco}, followed by bidirectional partial inference \cite{tan2021editspeech}, context-aware mask prediction \cite{wang2022campnet}, and explicit acoustic and prosodic consistency modeling \cite{liu2023fluenteditor}. Recent methods operate primarily on neural codec tokens or large generative speech models, including codec language models \cite{wang2024speechx}, zero-shot codec-token infilling and the RealEdit benchmark \cite{peng2024voicecraft}, robust zero-shot editing \cite{wang2025ssr}, flow-matching-based infilling \cite{le2023voicebox}, masked speech modeling with acoustic diffusion \cite{cambara2024mapache}, and natural-language-instructed editing \cite{huang2024instructspeech}.

\section{Preliminaries}
\label{sec:preliminaries}

\subsection{Discrete Diffusion Models}
\label{subsec:ddm_prelim}

Continuous diffusion models (CDMs)~\cite{ho2020denoising, song2020score} have become a powerful framework for generative modeling, with especially strong results in image generation~\cite{dhariwal2021diffusion}. However, applying diffusion directly to raw audio is challenging due to the high dimensionality and temporal resolution of waveform signals. A common way to reduce this complexity is to represent audio using compact discrete tokens obtained from a learned quantized CB~\cite{defossez2022high}. DDMs, which define the diffusion process over such token spaces, have recently shown strong results in natural language generation~\cite{lou2310discrete, nie2025large, sahoo2024simple}.
As in continuous diffusion, DDMs consist of a forward process that gradually corrupts a clean sample \(x_0\) into \(x_T\), and a learned reverse process that progressively reconstructs the original data. In absorbing-state DDMs, this corruption specifically replaces tokens with a mask state, such that \(x_T\) is fully masked.

\paragraph{Forward process.}
Let \(x_0 \sim p_{\text{data}}\) denote tokenized audio over a finite alphabet \(\mathcal{X} = \{1, \ldots, N\}\). The forward noising process is specified through a time-dependent transition rate matrix \(Q_t\), which determines the probability of moving between token states. In continuous time, this transition behavior is written as
\begin{equation}
\mathbb{P}(x_{t+\Delta t} = y \mid x_t = x)
=
\delta_{xy} + Q_t(y, x)\Delta t + \mathcal{O}(\Delta t^2),
\end{equation}
where \(Q_t(y,x)\) is the instantaneous rate of transitioning from state \(x\) to state \(y\) at time \(t\).

Following prior work~\cite{lou2310discrete}, the discrete forward diffusion process can be defined using a token-level transition rate matrix \(Q_{\text{tok}} \in \mathbb{R}^{n \times n}\), where \(n\) is the vocabulary size. For a single initial token \(x_0\), the marginal transition distribution at time \(t\) is given by
\begin{equation}
    p^{\text{tok}}_{t|0}(\cdot \mid x_0)
    =
    \exp\!\big(\bar\sigma(t) Q_{\text{tok}}\big),
    \label{eq:forward_marginal}
\end{equation}
where \(\bar\sigma(t)=\int_0^t \sigma(s)\,ds\) is the accumulated noise level. In the absorbing-mask formulation, \(Q_{\text{tok}}\) is chosen such that tokens eventually transition into a terminal \texttt{[MASK]} state. Under this design, the corruption process has a simple interpretation: each token either remains unchanged with probability \(e^{-\bar\sigma(t)}\), or is replaced by the \texttt{[MASK]} token with probability \(1-e^{-\bar\sigma(t)}\).

\paragraph{Reverse process.}
The reverse dynamics are also governed by a transition rate matrix. Following the time-reversal formulation of Markov processes~\cite{sun2022score, kelly1981reversibility}, the reverse transition rates are
\begin{equation}
\bar{Q}_t(y,x)
=
\frac{p_t(y)}{p_t(x)} Q_t(x,y),
\quad
\bar{Q}_t(x,x)
=
-\sum_{y \neq x} \bar{Q}_t(y,x).
\end{equation}
Thus, simulating the reverse process requires estimating the probability ratio \(\frac{p_t(y)}{p_t(x)}\). This ratio is commonly referred to as the \emph{concrete score}~\cite{lou2310discrete, meng2022concrete}, and can be parameterized by a neural network:
\begin{equation}
s_\theta(x,t)
\approx
\left[
\frac{p_t(y)}{p_t(x)}
\right]_{y \in \mathcal{X},\, y \neq x}.
\label{eq:concrete_score}
\end{equation}

\paragraph{Training.}
The neural score function is trained using the Diffusion Weighted Denoising Score Entropy (DWDSE) objective~\cite{lou2310discrete}. The loss is defined as

\begin{equation}
\small
\begin{aligned}
\mathcal{L}_{\text{DWDSE}}
&=
\int_0^T
\mathbb{E}_{\substack{
x_t \sim p_{t|0}(\cdot \mid x_0)
}}
\Bigg[
\sum_{y \neq x_t}
Q_t(x_t,y)
\Bigg(
s_\theta(x_t,t)_y
\\
&\qquad
-
\frac{p_{t|0}(y \mid x_0)}
     {p_{t|0}(x_t \mid x_0)}
\log s_\theta(x_t,t)_y
+
C
\Bigg)
\Bigg]\,dt .
\end{aligned}
\label{eq:dwdse-loss}
\end{equation}
where

\[
C = K\!\left(
\frac{p_{t|0}(y \mid x_0)}
     {p_{t|0}(x_t \mid x_0)}
\right),
\qquad
K(a) := a \log a - a.
\]
After training, reverse sampling is performed token by token. Given the current sample \(\mathbf{x}_t\), each token \(x_{t-\Delta t}^i\) is sampled according to

\begin{equation}
\begin{aligned}
p(x_{t-\Delta t}^i \mid x_t^i)
&=
\delta_{x_t^i}\!\left(x_{t-\Delta t}^i\right)
\\
&\quad+
\Delta t\,
Q_t^{\mathrm{tok}}\!\left(x_t^i,x_{t-\Delta t}^i\right)
s_\theta(\mathbf{x}_t,t)_{i,x_{t-\Delta t}^i}.
\end{aligned}
\label{eq:token_reverse}
\end{equation}
Repeating this reverse update from the fully corrupted state \(x_T\) to \(x_0\) yields samples from the learned generative model.

\subsection{Residual Vector Quantization}
Neural audio codecs such as EnCodec~\cite{defossez2022high} represent speech using RVQ. The encoder maps a waveform to $L$ latent frames, which are quantized sequentially by $K$ CBs. The first CB captures coarse acoustic information, while each subsequent CB quantizes the residual left by the preceding levels, progressively refining the representation.

We organize the resulting discrete representation as a $K \times L$ token grid, where the $k$-th row
\begin{equation}
\small
\mathbf{x}_0^{(k)}
=
\bigl(
x_0^{(k),1},\ldots,x_0^{(k),L}
\bigr)
\in
\bigl(\mathcal{X}^{(k)}\bigr)^L
\end{equation}
is the length-$L$ token sequence produced by CB $k$, and $x_0^{(k),i}$
denotes the token selected from CB $k$ at frame $i$. Thus,
$\mathbf{X}_0$ contains $L$ temporal frames and $K\times L$ discrete tokens, with
its rows following the inherent coarse-to-fine RVQ hierarchy.

\section{Speech Inpainting and Editing via Discrete Diffusion}
\label{sec:our_method}

\begin{figure*}[t]
    \centering
    \includegraphics[width=\textwidth]{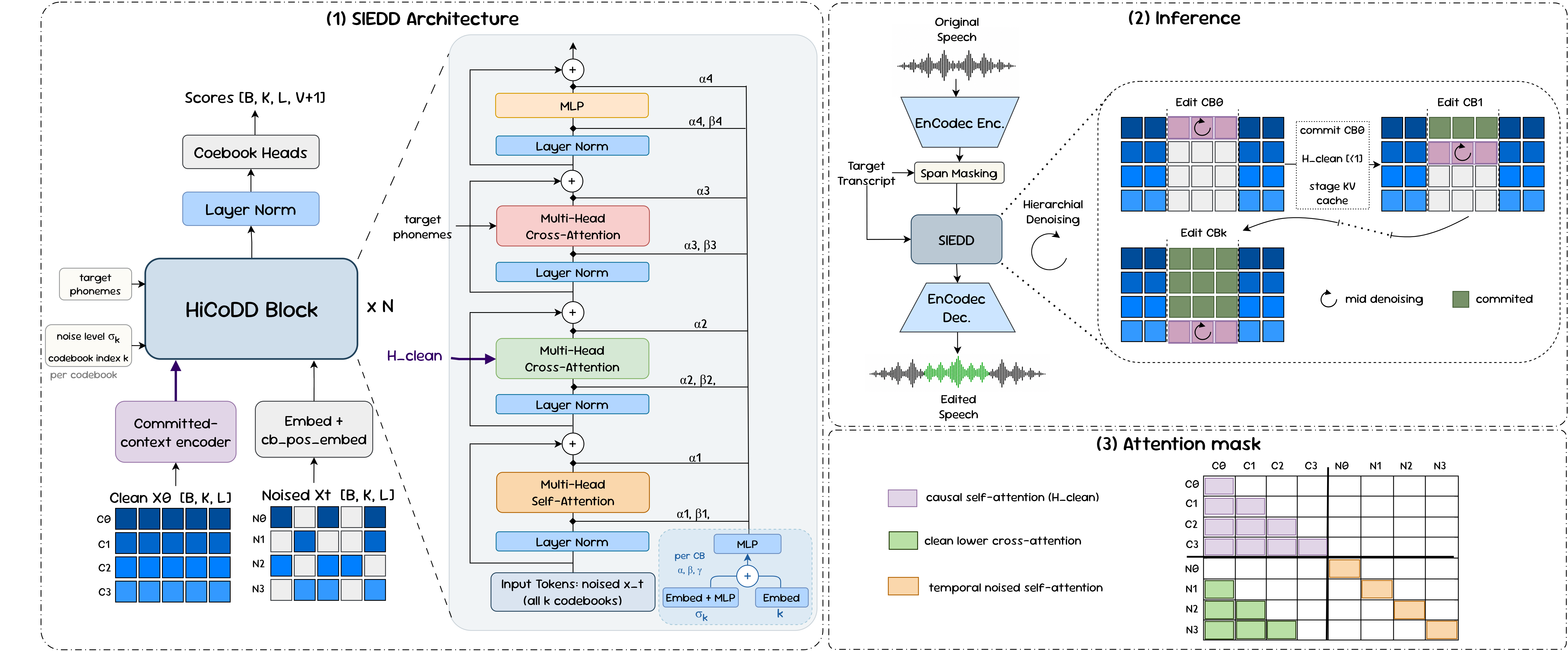}
    \caption{\textbf{Overview of the proposed architecture and inference procedure.}
    \textbf{(1) Architecture:} for each target CB $k$, the noised tokens are denoised by the diffusion decoder while the clean lower CBs $<k$ are encoded as committed acoustic context; each block combines temporal self-attention with cross-attention to the lower-CB and phoneme representations.\textbf{(2) Inference:} CBs are generated coarse to fine; at each stage only the edited region is denoised and then committed to the context of the next stage. The final tokens are decoded into the edited waveform. \textbf{(3) Attention mask:} the block-triangular pattern enforces the coarse-to-fine RVQ dependency and blocks access to the clean target CB, avoiding leakage. }
    \label{fig:system_overview}
\end{figure*}

We introduce SIEDD (see Figure~\ref{fig:system_overview}), a DDM framework for text-guided speech inpainting and editing. Given an input utterance that is either partially masked for inpainting or retained as fixed context for editing, together with a target text sequence, the model generates the codec tokens of the edited region while preserving the surrounding speech. The preliminaries describe absorbing-state discrete diffusion for a \emph{single} token sequence over a single alphabet $\mathcal{X}$. However, neural audio codecs such as EnCodec represent speech using a hierarchy of RVQ CBs with an inherent coarse-to-fine dependency. The central challenge is therefore to extend the single-sequence diffusion process including its score parameterization, training objective, and reverse sampler to this hierarchical setting without discarding the structure of RVQ. To address this challenge, we introduce \textbf{HiCoDD}. We further incorporate text conditioning into the score network, adapt guidance to focus on the edited span, and employ duration prediction to determine the length of the generated region.

\subsection{Hierarchical Multi-Codebook Discrete Diffusion}
\label{subsec:hier_multicodec}
Building on the RVQ representation introduced in the preliminaries, we augment each CB alphabet $\mathcal{X}^{(k)}$ with a dedicated \texttt{[MASK]} state and apply the absorbing corruption process at the level of an individual CB. Because each higher CB encodes a residual refinement of the lower CBs, the representation is inherently hierarchical: $\mathbf{x}^{(1)}_0$ captures the coarse acoustic structure, while each $\mathbf{x}^{(k)}_0$, $k>1$, is meaningful only in relation to the preceding levels.

\paragraph{Per-CB forward process.}
We apply the absorbing corruption of the preliminaries independently to each CB. Reusing the token-level rate matrix $Q_{\text{tok}}$ and accumulated noise schedule $\bar\sigma(t)$, the marginal corruption of CB $k$ is
\begin{equation}
    p^{(k)}_{t\mid0}\big(\cdot \mid x^{(k)}_0\big)
    =
    \exp\!\big(\bar\sigma(t)\,Q_{\text{tok}}\big).
\label{eq:per_cb_forward}
\end{equation}
The diffusion time is sampled independently for each CB, $t^{(k)}\!\sim\!\mathcal{U}(0,1]$, rather than shared across levels, so each utterance is seen at $K$ decorrelated noise levels.

\paragraph{Hierarchical reverse process.}
AR-based speech-editing methods exploit the RVQ ordering directly by predicting tokens in CB order. 
A naive application of discrete diffusion, in contrast, would flatten all CBs into
a single length-$K\times L$ sequence over the union alphabet $\bigcup_k \mathcal{X}^{(k)}$ and denoise them jointly
with one transition matrix $Q_{\text{tok}}$, thereby
recovering precisely the single-sequence process of the
preliminaries. Such a process, however, is
\emph{symmetric}: it treats all CBs identically and ignores the coarse-to-fine
dependency introduced by RVQ. We therefore keep the per-CB corruption defined above, but replace the single
generative distribution induced by the reverse process with a hierarchical
factorization inspired by~\cite{arriola2025block}:

\begin{equation}
\label{eq:hier_reverse}
p_\theta(\mathbf{X}\mid\mathbf{P})
=
p_\theta\!\big(\mathbf{x}^{(1)}\mid\mathbf{P}\big)
\prod_{k=2}^{K}
p_\theta\!\left(
\mathbf{x}^{(k)}
\mid
\mathbf{x}^{(<k)},\mathbf{P}
\right),
\end{equation}

where $\mathbf{x}^{(<k)}$ denotes the already generated lower CBs, $\mathbf{P}$ is the phoneme-level text representation, and each $p_\theta$ factor is induced by a reverse-diffusion chain parameterized by the concrete score $s_\theta$. Equation~\eqref{eq:hier_reverse} is therefore the multi-CB analogue of Eq.~\eqref{eq:token_reverse}, with the product enforcing that each CB refines an already established acoustic representation rather than reconstructing all levels independently.

\paragraph{Conditional concrete score.}
The single-sequence score of Eq.~\eqref{eq:concrete_score} approximates the
\emph{unconditional} probability ratio $p_t(y)/p_t(x)$. In the hierarchical
setting each CB is generated conditionally on its committed lower CBs and on the
target phoneme sequence, so we parameterize a \emph{conditional} concrete score. For CB $k$ at
position $i, y\neq x^{(k),i}_t$:
\begin{equation}
\small
s_\theta\!\big(\mathbf{x}^{(k)}_t,t \mid \mathbf{x}^{(<k)}_0,\mathbf{P}\big)_{i,y}
\;\approx\;
\frac{p_t\!\big(y \mid \mathbf{x}^{(<k)}_0,\mathbf{P}\big)}
     {p_t\!\big(x^{(k),i}_t \mid \mathbf{x}^{(<k)}_0,\mathbf{P}\big)}.
\label{eq:cond_concrete_score}
\end{equation}
Setting $K=1$ and dropping the conditioning recovers
Eq.~\eqref{eq:concrete_score} exactly; the hierarchy therefore enters the model
only through the conditioning arguments of a single, shared score network.

\paragraph{Hierarchical score-entropy training.}
Because every factor in Eq.~\eqref{eq:hier_reverse} is an absorbing diffusion
process with the marginal~\eqref{eq:per_cb_forward}, each level is trained with
the DWDSE objective of Eq.~\eqref{eq:dwdse-loss}, substituting the conditional
score of Eq.~\eqref{eq:cond_concrete_score} and summing over the positions of the CB;
denote this per-CB loss $\mathcal{L}^{(k)}$. The total objective averages over the
valid RVQ levels, $\mathcal{L}=\frac{1}{K}\sum_{k=1}^{K}\mathcal{L}^{(k)}$.
 
A key aspect of this scheme is that the model always conditions on the
\emph{ground-truth} lower CBs $\mathbf{x}^{(<k)}_0$, while the decoder for CB $k$
sees only the noised $\mathbf{x}^{(k)}_t$. This gives each refinement level the
exact clean context it will have after commitment at inference, rather than an approximation formed from partially noised lower CBs. The cross-CB attention is
restricted to the strictly lower CBs (the block-triangular mask in
Figure~\ref{fig:system_overview}~(3)), preventing the clean target CB from
reaching its own prediction.

\subsection{Hierarchical Diffusion Transformer}
\label{subsec:cross_att}
The score network realizing Eq.~\eqref{eq:cond_concrete_score} is a Diffusion
Transformer (DiT)~\cite{peebles2023scalable,vaswani2017attention} adapted to the
absorbing-state process of SEDD, organized around the separation between the
committed context and the CB currently being denoised.

\paragraph{Committed-context encoder.}
At stage $k$, the clean lower CBs $\mathbf{x}^{(<k)}_0$ are
treated as fixed context. A lightweight CB-causal encoder
contextualizes these tokens across time and CB levels, producing
a time-aware representation of their joint committed acoustic
state. Compared with attending directly to raw lower-CB
embeddings, this relieves the decoder from repeatedly inferring
their temporal alignment and residual composition at every
denoising step. The resulting representations supply the
lower-CB conditioning of Eq.~\eqref{eq:cond_concrete_score},
with the block-triangular visibility pattern illustrated in
Figure~\ref{fig:system_overview}~(3).

\paragraph{Denoising decoder.}
The noised tokens of the current CB form the main input sequence, and the
diffusion time $t$ modulates the network through adaptive layer normalization,
exactly mirroring how $t$ enters the score in Eq.~\eqref{eq:concrete_score}. To
supply the phoneme conditioning $\mathbf{P}$, we encode the target phoneme
sequence with a pretrained XPhoneBERT encoder~\cite{nguyen2023xphonebert},
project the resulting representations to the DiT hidden dimension, and inject them
through cross-attention. The committed lower CBs enter through a second
cross-attention stream over the context encoder's output, so that a single network jointly consumes both conditioning arguments of
Eq.~\eqref{eq:cond_concrete_score} (see Figure~\ref{fig:system_overview} (1)).
 
\subsection{Ordered Coarse-to-Fine Inference}
\label{subsec:inference}
Sampling follows the factorization of Eq.~\eqref{eq:hier_reverse}: CBs are
generated one stage at a time, from coarse to fine. Within stage $k$ we run the
single-CB reverse process, which is the conditional counterpart of the token
update in Eq.~\eqref{eq:token_reverse}. Given the current noised sequence
$\mathbf{x}^{(k)}_t$ and the committed context, each token is updated by

\begin{equation}
\small
\begin{aligned}
&p\!\left(
x_{t-\Delta t}^{(k),i}
\mid
\mathbf{x}_t^{(k)},\mathbf{x}_0^{(<k)},\mathbf{P}
\right)
=
\delta_{x_t^{(k),i}}
\!\left(x_{t-\Delta t}^{(k),i}\right)
\\
&\quad+
\Delta t\,
Q_t^{\mathrm{tok}}\!\left(
x_t^{(k),i},x_{t-\Delta t}^{(k),i}
\right)
s_\theta\!\left(
\mathbf{x}_t^{(k)},t
\mid
\mathbf{x}_0^{(<k)},\mathbf{P}
\right)_{i,x_{t-\Delta t}^{(k),i}} .
\end{aligned}
\label{eq:hier_token_reverse}
\end{equation}

Iterating Eq.~\eqref{eq:hier_token_reverse} from the fully masked state to
$t=0$ yields $\mathbf{x}^{(k)}_0$; the CB is then committed and appended to the
clean context for stage $k{+}1$.

\subsection{Localized Classifier-Free Guidance}
\label{subsec:cfg}
CFG~\cite{ho2022classifier} improves conditional generation by combining conditional and unconditional predictions. In speech
editing, accurate generation requires both adherence to the requested phoneme
sequence and preservation of the acoustic and linguistic context surrounding the
edited region. Since the network predicts categorical log-scores rather than
continuous noise, we apply guidance directly in log-score space on the
conditional score of Eq.~\eqref{eq:cond_concrete_score}:
\begin{equation}
\log s_{\mathrm{cfg}} = \alpha \log s_{\mathrm{cond}} + (1-\alpha)\log s_{\mathrm{neg}},
\label{eq:localized_cfg}
\end{equation}

where $\alpha$ is the guidance strength and $s_{\mathrm{neg}}$ is a negatively-conditioned concrete score. For the negative branch, we build on the random-phoneme conditioning strategy of~\cite{wang2025ssr}, which provides a stronger content-specific contrast than an empty prompt. We extend this idea into a localized CFG formulation tailored to HiCoDD. Rather than contrasting AR token logits, our guidance operates directly on the concrete scores in log-ratio space, the quantities that parameterize the reverse denoising process in Eq.~\eqref{eq:cond_concrete_score}. Moreover, we construct the negative branch by randomizing only the phonemes within the edited span, while keeping the surrounding phonemes and committed lower-CB representations identical to those of the conditional branch. This design isolates the guidance signal to the requested linguistic modification: it strengthens adherence to the target phonemes without suppressing the boundary cues required for coarticulation and seamless integration with the unedited speech. In inpainting the transcript is unchanged and the gap is specified in time rather than as a span of words, so no distinguished subsequence of phonemes exists to contrast against; we therefore randomize the full phoneme sequence for the negative branch. 

\subsection{Localized Generation: Inpainting and Editing}
\label{subsec:gap}
Both tasks reduce to the same localized instance of
Eq.~\eqref{eq:hier_token_reverse}: the tokens outside the target span are held
fixed as clean context, and only the tokens inside the span are initialized to
\texttt{[MASK]} and denoised. The two tasks differ solely in how the length of
that span is determined. In \emph{inpainting}, the duration of the missing segment is known and maps
directly, through the tokenizer frame rate, to the number of acoustic tokens to
mask and regenerate. In \emph{text-based editing} particularly for insertion
and substitution the duration of the edited segment is not known in advance,
since speaking rates vary across speakers and the mapping from phonemes to
acoustic-token durations is not fixed. To resolve this, we adopt a duration
predictor inspired by FastSpeech~\cite{ren2020fastspeech}. Further architectural and
training details are provided in the supplementary material.

\begin{figure*}[t]
  \centering
  \includegraphics[width=\textwidth]{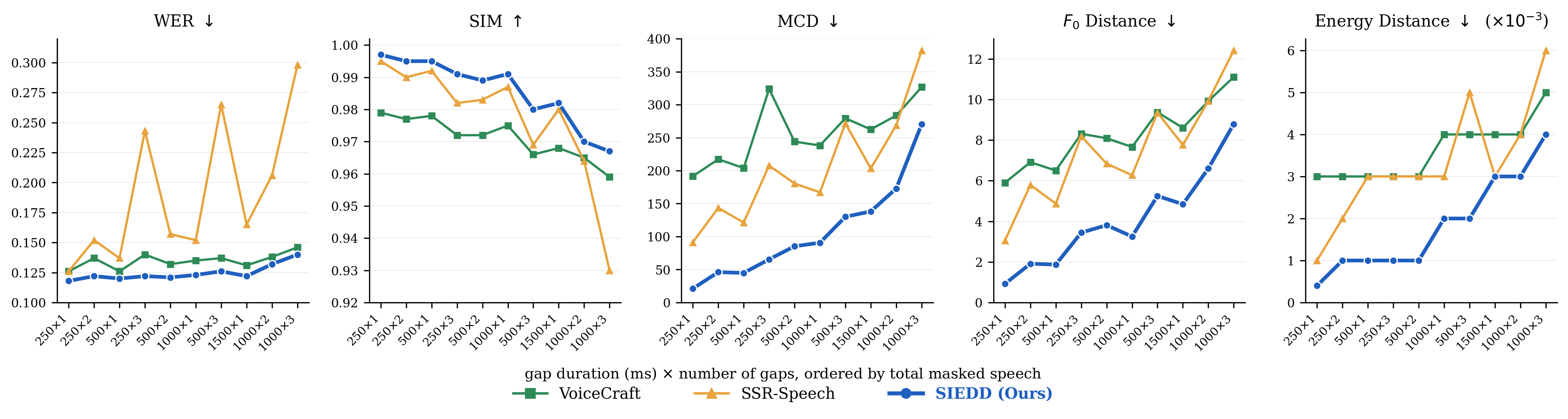}
  \caption{Speech inpainting results across masked-span configurations, ordered by
  total masked speech. Lines show the mean. Higher is better for SIM, lower for all other metrics.}
  \label{fig:speech_inpainting_results}
\end{figure*}
\begin{table*}[t]
\centering
\scriptsize

\begin{tabular}{lcccccc}
\toprule
\textbf{Method}
& \textbf{WER $\downarrow$}
& \textbf{SIM $\uparrow$}
& \textbf{MCD $\downarrow$}
& \textbf{$F_0$ Dist. $\downarrow$}
& \textbf{Energy Dist. $\downarrow$}
& \textbf{UTMOS $\uparrow$} \\
\midrule
TTS Baseline (MMS) & 0.147$\pm$0.13 & - & - & - & - &  \textbf{4.11$\pm$0.19} \\
\midrule
VoiceCraft   & 0.124$\pm$0.11 & 0.97$\pm$0.02 & 392.25$\pm$134.9 & 9.94$\pm$6.9 & 0.005$\pm$0.004 & 3.47$\pm$0.63 \\
SSR-Speech   & 0.146$\pm$0.12 & 0.97$\pm$0.02 & 308.3$\pm$156.1 & \textbf{8.17$\pm$7.79} & 0.005$\pm$0.004 & 3.45$\pm$0.67 \\
\textbf{SIEDD (Ours)}  & \textbf{0.121$\pm$0.11} & \textbf{0.98$\pm$0.03} & \textbf{270.0$\pm$95.1} & 8.57$\pm$6.9 & \textbf{0.005$\pm$0.003} & 3.44$\pm$0.59 \\
\bottomrule
\end{tabular}
\caption{Speech editing results on RealEdit. Results are reported as mean $\pm$ standard deviation. Higher is better for SIM and UTMOS, while lower is better for all other metrics. In \textbf{bold} marked the best result.}
\label{tab:realedit_results}
\end{table*}

\section{Experiments}
\label{sec:experiments}

\subsection{Metrics}
\label{subsec:metrics}

We evaluate intelligibility using WER from Whisper medium-en~\cite{radford2022whisper}; speaker preservation using cosine similarity between WavLM~\cite{chen2022wavlm} embeddings; spectral fidelity using 13-dimensional MFCC-based MCD; pitch and energy consistency using the mean absolute differences between $F_0$ and RMS-energy contours, respectively; and predicted perceptual quality using the UTMOS22 strong learner~\cite{saeki2022utmos}. All frame-level acoustic metrics are computed after DTW alignment and normalized by the alignment-path length. For $F_0$, we use pYIN over 80--600 Hz and assign zero to unvoiced frames.

\subsection{Datasets.}
For both speech-editing and speech-inpainting evaluation, we use RealEdit~\cite{peng2024voicecraft}, a manually curated benchmark of realistic examples drawn from LibriTTS~\cite{zen2019libritts} and GigaSpeech~\cite{chen2021gigaspeech} YouTube recordings\footnote{Spotify podcasts are not publicly available anymore, so it is excluded from our evaluations.}. The benchmark covers \textit{insertion, deletion, substitution}, and multi-span edits across diverse speakers, accents, recording conditions, background sounds, audio qualities, and sampling rates.

\subsection{Evaluation.}
We evaluate SIEDD on the RealEdit benchmark following the evaluation protocol of \cite{peng2024voicecraft}, using its audio utterances and corresponding editing instructions, which cover diverse recording conditions and edit types. For speech inpainting, we use the same RealEdit utterance list. We construct uniformly spaced masked regions of varying lengths, with up to three masked spans per utterance. We generate each utterance five times with different fixed seeds and report the average for each utterance.

\subsection{Baselines}
\label{subsec:baselines}

We evaluate our method on both text-based speech editing and speech inpainting, comparing it with several recent approaches. \textbf{VoiceCraft}~\cite{peng2024voicecraft} is a neural codec language model that performs speech editing through token infilling, while \textbf{SSR-Speech}~\cite{wang2025ssr} is a zero-shot framework for text-based speech editing and synthesis based on discrete speech tokens. We additionally include \textbf{MMS}~\cite{pratap2023mms} as a zero-shot TTS baseline that synthesizes the target text directly. VoiceCraft, SSR-Speech, and SIEDD all use a 16-kHz EnCodec tokenizer trained on GigaSpeech XL, and the models themselves are also trained on the same dataset, yielding a controlled comparison that primarily isolates the generative formulation: AR modeling versus our discrete diffusion approach.

\subsection{Training.}
Following the baselines, we train SIEDD for 550{,}000 steps on two NVIDIA RTX A6000 GPUs, requiring approximately one week on GigaSpeech XL. Learning rate of $10^{-4}$ and AdamW as optimizer. The duration predictor was trained on LibriTTS train-clean. Full training and hyperparameter details are provided in the supplementary material.

\section{Results}

\paragraph{Speech Inpainting.}
Figure~\ref{fig:speech_inpainting_results} shows that SIEDD consistently outperforms the baselines across the evaluated settings, with particularly large gains for short and medium gaps. For a single $250$~ms gap, it reduces MCD to $21.1$, compared with $91.3$ for SSR-Speech and $191.5$ for VoiceCraft, while also achieving the best WER, SIM, $F_0$-distance, and energy-distance scores. Although performance degrades as the number and duration of the gaps increase, SIEDD remains the strongest method in nearly all configurations, demonstrating effective reconstruction of both linguistic content and local acoustic characteristics. Notably, the AR baselines exhibit sharp WER increases in the multi-gap settings, whereas SIEDD remains substantially more stable. We hypothesize that sequential AR generation is more susceptible to error accumulation: inaccuracies introduced while reconstructing one region may affect the context used to generate subsequent regions, particularly when several disjoint gaps must be completed. In contrast, the diffusion-based formulation repeatedly refines the masked regions while conditioning on the available context, reducing irreversible propagation of early decoding errors. Full numerical results are provided in the supplementary material.

\paragraph{Speech Editing.}
As shown in Table~\ref{tab:realedit_results}, SIEDD achieves the best overall speech-editing performance. It obtains the lowest WER ($0.121$), the highest speaker similarity ($0.98$), the lowest MCD ($270.0$), and the lowest energy distance. Although the TTS baseline obtains the highest UTMOS score, this is unsurprising, as it synthesizes the target speech independently rather than adapting it to the speaker, acoustic conditions, and background of the original recording. In comparison, the original unedited audio receives a UTMOS score of $3.56$, and all three speech-editing methods obtain similar scores, ranging from $3.44$ to $3.47$. This suggests that their outputs remain close to the perceptual quality of the original recordings. Overall, SIEDD provides the strongest balance between content accuracy, speaker preservation, acoustic consistency, and perceptual quality.

\section{Ablation Study}
\label{sec:ablation_study}

Table~\ref{tab:realedit_ablation} ablates the four main design choices of SIEDD on
RealEdit: the tokenizer representation, the CB factorization, the duration predictor
(DP), and CFG. All variants use 512 denoising steps, under the same evaluation setup. Each experiment is isolating a different aspect of our architectural choices, and are identical unless stated otherwise.

\paragraph{Single- vs.\ multi-CB tokenization.}
The single-CB baseline replaces EnCodec with WavTokenizer~\cite{ji2024wavtokenizer} and \textit{HiCoDD} with a vanilla DiT backbone augmented with target-phoneme conditioning through cross-attention. Because it has no CB axis, it does not include cross-CB modeling. Moving to the multi-CB representation improves WER ($0.186 \rightarrow 0.139$) and SIM ($0.95 \rightarrow 0.97$), indicating that the coarse-to-fine residual structure of RVQ better preserves linguistic content and speaker identity than a single heavily compressed stream in our discrete diffusion framework.

\paragraph{Joint vs.\ hierarchical CB modeling.}                                                           \emph{Multi (joint)} is the non-hierarchical counterpart of Eq.~\eqref{eq:hier_reverse}: all $K$ CBs are corrupted and denoised \emph{simultaneously}, so the model estimates a single joint score over the full $K{\times}L$ grid instead of the ordered factorization. It retains the per-CB embedding tables and per-CB output heads, but replaces HiCoDD's clean-context encoder and lower-CB cross-attention with \emph{symmetric} cross-CB attention: each block alternates within-CB temporal self-attention with attention across the CB axis, so every CB attends to every other at the same frame, with no coarse-to-fine ordering and no separation between clean context and the CB being denoised. Phoneme cross-attention, tokenizer, and training data are unchanged. Consequently a fine CB may be committed before the coarse CB it refines has been resolved. At the same CFG strength, hierarchical factorization improves WER ($0.152 \rightarrow 0.121$).

\paragraph{Duration predictor and CFG.}
\emph{no DP} variant uses a phoneme-rate heuristic, $|\text{ph}_{\text{new}}| / r \times
f$, with the speaker rate $r$ calibrated from the MFA alignment of the original utterance
and $f$ the tokenizer frame rate; it is linear in phoneme count and blind to phoneme
identity and edit position, rather than using the learned predictor. Removing the DP degrades WER ($0.121 \rightarrow 0.136$) and SIM ($0.98 \rightarrow 0.97$). Similarly, by removing localized CFG degrades WER ($0.121 \rightarrow 0.139$) and SIM ($0.98 \rightarrow 0.97$), confirming that the log-score contrast of Eq.~\ref{eq:localized_cfg} strengthens adherence to the requested phonemes.

\begin{table}[t]
\centering
\scriptsize

\setlength{\tabcolsep}{5pt}
\begin{tabular}{lcccc}
\toprule
\textbf{Tokenizer}
& \textbf{Codec}
& \textbf{CFG}
& \textbf{WER $\downarrow$}
& \textbf{SIM $\uparrow$} \\
\midrule

WavTokenizer (24 kHz)
& Single
& \checkmark
& 0.186 $\pm$0.16
& 0.95$\pm$0.3 \\

EnCodec (16 kHz)
& Multi
& $\times$
& 0.139$\pm$0.12 
& 0.97$\pm$0.02 \\

EnCodec (16 kHz)
& Multi (joint)
& \checkmark
& 0.152$\pm$0.12
& 0.97$\pm$0.03 \\

EnCodec (16 kHz)
& Multi (without DP)
& \checkmark
&  0.136$\pm$0.12
& 0.97$\pm$0.02 \\

\midrule
EnCodec (16 kHz)
& Multi (SIEDD)
& \checkmark
& \textbf{0.121$\pm$0.11}
& \textbf{0.98$\pm$0.03} \\

\bottomrule

\end{tabular}
\caption{Ablation study on RealEdit speech editing. A \checkmark\ indicates that CFG is used. DP stands for duration predictor. All the results are on SIEDD with 512 denoising steps. Unless stated otherwise DP was used.}
\label{tab:realedit_ablation}
\end{table}

\section{Conclusion}

We introduced SIEDD, a discrete diffusion framework for speech text-guided inpainting and  editing over multi-CB neural codec representations. Its core architecture, HiCoDD, follows the coarse-to-fine structure of RVQ tokenizers by treating previously generated CBs as clean, committed acoustic context while applying diffusion only to the current refinement CB. This enables leakage-free joint training while remaining consistent with sequential inference.

SIEDD further combines phoneme-level conditioning, duration prediction, and span-localized CFG adapted to categorical diffusion. These components strengthen adherence to the target content while preserving the surrounding linguistic and acoustic context, supporting both fixed-duration inpainting and variable-duration editing.

Experiments on RealEdit show that SIEDD achieves a strong balance of intelligibility, speaker preservation, and acoustic consistency, outperforming the evaluated AR baselines across nearly all inpainting settings and achieving the best overall editing results. Ablations confirm that these gains arise from both hierarchy-aware multi-CB modeling and CFG. Overall, the results establish SIEDD as a promising alternative to AR generation for context-preserving speech modification, with future directions including faster sampling, improved duration modeling, multilingual speech, and longer or more general edits.

\bibliographystyle{plainnat}   
\bibliography{ref}

\section{Reproducibility}
Full code and configuration files are released with the paper. This section
summarizes the hyperparameters used for our main method, HiCoDD
(\texttt{configs/config\_encodec\_hier.yaml}).

\subsection{Model Architecture}

The main architectural configuration of HiCoDD is summarized in
Table~\ref{tab:repro-model}.

\subsection{Diffusion Process}

The forward-process graph and noise-schedule configuration are reported in
Table~\ref{tab:repro-diffusion}.

\subsection{Training Hyperparameters}

The settings used to train HiCoDD are provided in
Table~\ref{tab:repro-train}.

\subsection{Inference and Sampling Hyperparameters}

HiCoDD samples codebooks in an ordered, coarse-to-fine manner: each codebook
is fully denoised before conditioning the generation of the next codebook
through the clean encoder. The sampling hyperparameters are summarized in
Table~\ref{tab:repro-sampling}.

\section{Token-Count Duration Prediction}
\label{sec:duration_predictor}

SIEDD performs speech editing directly in the discrete token space of an EnCodec codec. For substitution and insertion operations, the system must determine the number of codec frames allocated to the edited span before generation. A simple approach is to estimate this duration from the number of phonemes in the target text and the speaking rate of the source speaker:
\begin{equation}
\hat{N}_{\mathrm{rate}}
\frac{|\phi_{\mathrm{new}}|}{r_{\mathrm{spk}}}
\cdot f_{\mathrm{codec}},
\label{eq:rate_duration}
\end{equation}
where $|\phi_{\mathrm{new}}|$ denotes the number of phonemes in the edited span, $r_{\mathrm{spk}}$ is the estimated speaker phoneme rate, and $f_{\mathrm{codec}}=50$ is the EnCodec frame rate. Although inexpensive, this heuristic assumes that all phonemes contribute equally to duration and does not account for phoneme identity, span position, coarticulation, or local prosodic variation. It is particularly unreliable for insertions, for which no original acoustic span is available to provide a direct duration reference.

To obtain a more accurate estimate, we introduce a lightweight neural token-count predictor inspired by the convolutional duration predictor of FastSpeech~2~\citep{ren2020fastspeech}. The predictor receives the IPA phoneme sequence of the target span and outputs the expected number of EnCodec frames required to realize it.

\paragraph{Architecture.}
Each input phoneme is mapped to a $d$-dimensional embedding, with $d=128$, and augmented with a fixed sinusoidal positional encoding~\citep{vaswani2017attention}. The resulting sequence is processed by two convolutional blocks. Each block consists of a one-dimensional convolution with kernel size $3$, followed by a residual connection, layer normalization, ReLU activation, and dropout with probability $0.1$.

The encoded phoneme sequence is aggregated using masked mean pooling, such that padding positions do not contribute to the span representation. To account for duration variation that cannot be inferred from phoneme identity alone, the pooled representation is concatenated with three scalar conditioning variables:
\begin{equation}
\mathbf{c}
\left[
\log r_{\mathrm{spk}},
;
\rho_{\mathrm{pos}},
;
\log |\phi_{\mathrm{new}}|
\right],
\end{equation}
where $\rho_{\mathrm{pos}}\in[0,1]$ denotes the normalized location of the span center within the utterance. The combined representation is passed through a two-layer multilayer perceptron,
\begin{equation}
\mathbb{R}^{d+3}
\rightarrow
\mathbb{R}^{d/2}
\rightarrow
\mathbb{R},
\end{equation}
which predicts the log token count. The final integer-valued estimate is obtained as
\begin{equation}
\hat{N}
\operatorname{round}
\left(
\exp\left(g_{\theta}(\phi_{\mathrm{new}},\mathbf{c})\right)
\right).
\end{equation}
The predictor contains approximately $10^5$ parameters and therefore introduces negligible computational overhead relative to the diffusion backbone.

\paragraph{Training data.}
We construct span-level duration supervision from the LibriTTS corpus. Word-level timestamps are obtained using the Montreal Forced Aligner. For a contiguous span covering words $i$ through $j$, the target number of codec frames is defined as
\begin{equation}
N^{*}
\left(
t_{\mathrm{end},j}
t_{\mathrm{start},i}
\right)
f_{\mathrm{codec}},
\label{eq:duration_target}
\end{equation}
where $t_{\mathrm{start},i}$ and $t_{\mathrm{end},j}$ are the aligned start and end times of the span. We enumerate contiguous spans of up to eight words from each utterance, producing approximately $180$--$200$k training examples. The resulting dataset is divided into training and validation subsets using a $90/10$ split.

\paragraph{Training objective.}
Rather than directly minimizing absolute token-count error, we optimize a Huber loss over the predicted-to-reference duration ratio:
\begin{equation}
\mathcal{L}_{\mathrm{dur}}
\operatorname{Huber}_{\delta=0.3}
\left(
\frac{\hat{N}}{N^{*}},
1
\right).
\label{eq:duration_loss}
\end{equation}
Operating in relative-duration space prevents long spans from dominating the objective and assigns comparable importance to proportional errors across short and long edits. The Huber formulation additionally reduces sensitivity to noisy alignments and unusually long or short realizations.

The model is trained with AdamW using a learning rate of $10^{-3}$ and weight decay of $10^{-4}$. We apply cosine learning-rate annealing down to $10^{-5}$, gradient clipping with a maximum norm of $1.0$, and a batch size of $256$. Training proceeds for at most $50$ epochs, with early stopping after $10$ epochs without validation improvement. The checkpoint with the lowest validation mean absolute error in codec tokens is retained. We compare the learned predictor against the phoneme-rate heuristic in Eq.~\eqref{eq:rate_duration} to quantify the benefit of modeling phonetic and contextual duration variation.

\begin{table*}[t]
\centering
\scriptsize
\begin{tabular}{ll}
\toprule
Parameter & Value \\
\midrule
Backbone & DiT \\
Tokenizer & EnCodec \\
Hidden size & 1024 \\
Conditioning dim & 128 \\
Transformer blocks & 24 \\
Attention heads & 16 \\
Dropout & 0.1 \\
$\sigma$-scaling of final layer & True \\
Codebooks ($K$) & 4 \\
Clean-encoder depth & 1 block \\
Encoder fusion mode & causal (codebook-causal) \\
Per-codebook positional embedding & True \\
Cross-attention conditioning & True \\
adaLN global conditioning & False \\
Phoneme conditioning (XPhoneBERT) & True \\
Text positional embedding & True \\
Vocabulary size per codebook & 2048 (+1 absorbing/mask state) \\
\bottomrule
\end{tabular}
\caption{HiCoDD architecture.}
\label{tab:repro-model}
\end{table*}

\begin{table}[t]
\centering
\scriptsize
\begin{tabular}{ll}
\toprule
Parameter & Value \\
\midrule
Graph type & absorbing \\
Noise schedule & log-linear \\
$\sigma_{\min}$ & $1\times10^{-4}$ \\
$\sigma_{\max}$ & 20 \\
Masking mode & iid, independent per codebook \\
Per-codebook noise coupling & independent (no staircase) \\
\bottomrule
\end{tabular}
\caption{Forward process: graph and noise schedule.}
\label{tab:repro-diffusion}
\end{table}

\begin{table}[t]
\centering
\scriptsize
\begin{tabular}{ll}
\toprule
Parameter & Value \\
\midrule
Optimizer & AdamW \\
Learning rate & $1\times10^{-4}$ \\
LR warmup & 2{,}500 steps \\
Gradient clipping & 1.0 \\
GPUs & 2 \\
Per-GPU batch size & 8 \\
Gradient accumulation & 2 steps \\
EMA decay & 0.9999 \\
\bottomrule
\end{tabular}
\caption{Training and optimization hyperparameters.}
\label{tab:repro-train}
\end{table}

\begin{table}[t]
\centering
\small
\begin{tabular}{ll}
\toprule
Parameter & Value \\
\midrule
Mode & ordered (coarse $\to$ fine) \\
Steps & 512 \\
Temperature & 1.0 \\
Top-$k$ & 2 \\
CFG coefficient & 1.5 \\
Seed & 1 \\
\bottomrule
\end{tabular}
\caption{Sampling hyperparameters.}
\label{tab:repro-sampling}
\end{table}

\section{Full Inpainting Evaluations}
In Table~\ref{tab:speech_inpainting_results} showed the numerical results of the graphs in the main paper.

\begin{table*}[t]
\centering
\scriptsize
\setlength{\tabcolsep}{4pt}
\begin{tabular}{cclccccc}
\toprule
\textbf{Gap Duration}
& \textbf{\# Gaps}
& \textbf{Method}
& \textbf{WER $\downarrow$}
& \textbf{SIM $\uparrow$}
& \textbf{MCD $\downarrow$}
& \textbf{$F_0$ Dist. $\downarrow$}
& \textbf{Energy Dist. $\downarrow$}\\
\midrule

\multirow{12}{*}{\textbf{250 ms}}
& \multirow{4}{*}{1}
& VoiceCraft   & 0.126$\pm$0.11 & 0.979$\pm$0.02 & 191.5$\pm$35.0 & 5.90$\pm$4.2 & 0.003$\pm$0.001 \\
& & SSR-Speech   & 0.126$\pm$0.11 & 0.995$\pm$0.01 & 91.3$\pm$45.6 & 3.05$\pm$3.18 & 0.001$\pm$0.001 \\
& & \textbf{SIEDD (Ours)}  & \textbf{0.118$\pm$0.10} & \textbf{0.997$\pm$0.004} & \textbf{21.1$\pm$14.7} & \textbf{0.92$\pm$1.27} & \textbf{0.0004$\pm$0.001} \\
\cmidrule(lr){2-8}
& \multirow{4}{*}{2}
& VoiceCraft   & 0.137$\pm$0.12 & 0.977$\pm$0.02 & 217.2$\pm$54.7 & 6.91$\pm$4.54 & 0.003$\pm$0.002 \\
& & SSR-Speech   & 0.152$\pm$0.13 &  0.990$\pm$0.02 & 143.5$\pm$71.3 & 5.79$\pm$6.30 & 0.002$\pm$0.0020 \\
& & \textbf{SIEDD (Ours)}  & \textbf{0.122$\pm$0.11} & \textbf{0.995$\pm$0.01} & \textbf{46.07$\pm$31.7} & \textbf{1.91$\pm$2.27} & \textbf{0.001$\pm$0.001} \\

\cmidrule(lr){2-8}
& \multirow{4}{*}{3}
& VoiceCraft   & 0.140$\pm$0.12 & 0.972$\pm$0.03 & 323.89$\pm$47.6 & 8.31$\pm$7.53 & 0.003$\pm$0.002 \\
& & SSR-Speech   & 0.243$\pm$0.77 & 0.982$\pm$0.02 & 207.7$\pm$79.5 & 8.19$\pm$6.62 & 0.003$\pm$0.002 \\
& & \textbf{SIEDD (Ours)}  & \textbf{0.122$\pm$0.11} & \textbf{0.991$\pm$0.01} & \textbf{65.5$\pm$30.2} & \textbf{3.44$\pm$5.46} & \textbf{0.001$\pm$0.001} \\
\midrule

\multirow{12}{*}{\textbf{500 ms}}
& \multirow{4}{*}{1}
& VoiceCraft   & 0.126$\pm$0.11 & 0.978$\pm$0.02 & 204.0$\pm$38.6 & 6.50$\pm$5.3 & 0.003$\pm$0.002 \\
& & SSR-Speech   & 0.137$\pm$0.11 & 0.992$\pm$0.02 & 121.3$\pm$79.3 & 4.87$\pm$8.01 & 0.003$\pm$0.001 \\
& & \textbf{SIEDD (Ours)}  & \textbf{0.120$\pm$0.11} & \textbf{0.995$\pm$0.01} & \textbf{44.9$\pm$29.0} & \textbf{1.87$\pm$3.20} & \textbf{0.001$\pm$0.001} \\
\cmidrule(lr){2-8}
& \multirow{4}{*}{2}
& VoiceCraft   & 0.132$\pm$0.12 & 0.972$\pm$0.026 & 243.9$\pm$60.6 & 8.09$\pm$5.45 & 0.003$\pm$0.002 \\
& & SSR-Speech   & 0.157$\pm$0.125 & 0.983$\pm$0.02 & 180.5$\pm$77.6 & 6.84$\pm$5.14 & 0.003$\pm$0.002 \\
& & \textbf{SIEDD (Ours)}  & \textbf{0.121$\pm$0.11} & \textbf{0.989$\pm$0.01} & \textbf{85.3$\pm$44.1} & \textbf{3.81$\pm$4.20} & \textbf{0.001$\pm$0.001} \\
\cmidrule(lr){2-8}
& \multirow{4}{*}{3}
& VoiceCraft   & 0.137$\pm$0.12 & 0.966$\pm$0.03 & 279.4$\pm$72.0 & 9.37$\pm$7.74 & 0.004$\pm$0.002 \\
& & SSR-Speech   & 0.265$\pm$0.165 & 0.969$\pm$0.04 & 271.7$\pm$98.32 & 9.35$\pm$6.20 & 0.005$\pm$0.003 \\
& & \textbf{SIEDD (Ours)}  & \textbf{0.126$\pm$0.11} & \textbf{0.980$\pm$0.03} & \textbf{129.98$\pm$59.7} & \textbf{5.25$\pm$5.36} & \textbf{0.002$\pm$0.002} \\
\midrule

\multirow{12}{*}{\textbf{1000 ms}}
& \multirow{4}{*}{1}
& VoiceCraft   & 0.135$\pm$0.12 & 0.975$\pm$0.02 & 238.2$\pm$60.1 & 7.66$\pm$5.5 & 0.004$\pm$0.002\\
& & SSR-Speech   & 0.152$\pm$0.12 & 0.987$\pm$0.02 & 167.0$\pm$79.3 & 6.27$\pm$6.35 & 0.003$\pm$0.002 \\
& & \textbf{SIEDD (Ours)}  & \textbf{0.123$\pm$0.11} & \textbf{0.991$\pm$0.01} & \textbf{90.68$\pm$48.3} & \textbf{3.25$\pm$3.24} & \textbf{0.002$\pm$0.001} \\
\cmidrule(lr){2-8}
& \multirow{4}{*}{2}
& VoiceCraft   & 0.138$\pm$0.12 & 0.965$\pm$0.03 & 283.6$\pm$72.3 & 9.92$\pm$6.88 & 0.004$\pm$0.002 \\
& & SSR-Speech   & 0.206$\pm$0.15 & 0.964$\pm$0.05 & 268.8$\pm$109.8 & 9.93$\pm$8.39 & 0.004$\pm$0.003 \\
& & \textbf{SIEDD (Ours)}  & \textbf{0.132$\pm$0.12} & \textbf{0.970$\pm$0.04} & \textbf{172.9$\pm$82.7} & \textbf{6.60$\pm$6.16} & \textbf{0.003$\pm$0.002} \\
\cmidrule(lr){2-8}
& \multirow{4}{*}{3}
& VoiceCraft   & 0.146$\pm$0.12 & 0.959$\pm$0.04 & 326.6$\pm$71.10 & 11.11$\pm$6.75 & 0.005$\pm$0.003 \\
& & SSR-Speech   & 0.298$\pm$0.167 & 0.930$\pm$0.09 & 382.1$\pm$125.5 & 12.42$\pm$7.81 & 0.006$\pm$0.004 \\
& & \textbf{SIEDD (Ours)}  & \textbf{0.140$\pm$0.12} & \textbf{0.967$\pm$0.08} & \textbf{270.4$\pm$122.80} & \textbf{8.79$\pm$7.11} & \textbf{0.004$\pm$0.003} \\
\midrule

\multirow{3}{*}{\textbf{1500 ms}}
& \multirow{3}{*}{1}
& VoiceCraft
& 0.131$\pm$0.12
& 0.968$\pm$0.04
& 262.5$\pm$65.5
& 8.60$\pm$5.7
& 0.004$\pm$0.002 \\

& & SSR-Speech
& 0.165$\pm$0.13
& 0.980$\pm$0.02
& 203.3$\pm$95.6
& 7.76$\pm$6.80
& 0.003$\pm$0.003 \\

& & \textbf{SIEDD{} (Ours)}
& \textbf{0.122$\pm$0.10}
& \textbf{0.982$\pm$0.02}
& \textbf{138.2$\pm$62.8}
& \textbf{4.84$\pm$3.73}
& \textbf{0.003$\pm$0.002} \\
\bottomrule
\end{tabular}
\caption{Speech inpainting results for different masked-span durations and numbers of gaps. Results are reported as mean $\pm$ standard deviation. Each gap has the duration indicated in the first column. Higher is better for SIM, while lower is better for all other metrics. In \textbf{bold} marked the best result. Evaluation with gaps that were longer than the utterance itself were skipped.}
\label{tab:speech_inpainting_results}
\end{table*}

\end{document}